\documentclass[prb,aps,twocolumn,groupedaddress]{revtex4}
\usepackage{epsfig,latexsym,amssymb,amsmath,amsbsy,graphics,graphicx}

\def \beq{\begin{equation}}
\def \eeq{\end{equation}}
\def \beqarr{\begin{eqnarray}}
\def \eeqarr{\end{eqnarray}}

\begin{document}

\title{Spontaneous Symmetry Breaking and Quantum Hall Effect in Graphene}

\author{Kun Yang}

\affiliation{National High Magnetic Field Laboratory and Department
of Physics, Florida State University, Tallahassee, FL 32306, USA}

\date{\today}

\begin{abstract}

In this article we briefly review recent experimental and
theoretical work on quantum Hall effect in graphene, and argue that
some of the quantum Hall states exhibit spontaneous symmetry
breaking that is driven by electron-electron interaction. We will
also discuss how to experimentally determine the actual manner in
which symmetry breaking occurs, and detect the collective charge and
neutral excitations associated with symmetry breaking. Other issues
will also be briefly mentioned.
\end{abstract}


\maketitle

\section{Introduction}

Recent experimental work\cite{graphene_exp} has established graphene
as a new two-dimensional (2D) electron system with linear Dirac-like
energy-band dispersion. Shortly after the original work performed at
zero magnetic field, experimentalist applied a strong magnetic field
(around 10T) perpendicular to the graphene sheet, and observed a set
of integer quantum Hall plateaus on which the Hall conductance takes
quantized values\cite{novoselov,zhang}:
\begin{equation}
\sigma_{xy}=\nu e^2/h,
\end{equation}
with
\begin{equation}
\nu =4(n+1/2), \label{berry}
\end{equation}
where $n$ is an integer. This result, which was anticipated
theoretically\cite{zheng,gusynin,peres}, may be understood in the
following manner: the prefactor 4 reflects the two-fold spin and
two-fold valley degeneracy in the graphene band structure, while the
``shift" of 1/2 originates from the Berry phase due to the
pseudospin (or valley) precession when a massless (and thus chiral)
Dirac particle exercises cyclotron motion. This results thus
provides direct evidence for the Dirac or relativistic nature of the
charge carriers in graphene; it can also be understood from the
Landau level structure of Dirac particles (see Appendix A for a
pedagogical discussion). Historically, novel electromagnetic
response of fermions moving on a honeycomb lattice (which graphene
is an example of), including the quantum Hall effect, was considered
by Semenoff\cite{semenoff} and Haldane\cite{haldane}, and
considerations of quantum Hall effect of relativistic particles
dates back even further.\cite{ahm83,schakel} We also note that such
integer quantum Hall plateaus were found to persist to room
temperature\cite{boebinger}, paving the way for their application
some day.

When a even stronger magnetic field (up to 45 T) is applied, more
quantum Hall (QH) plateaus {\em not} in the sequence of Eq.
(\ref{berry}) have been observed\cite{zhang2}; these include $\nu=0,
\pm 1$, and $ \pm 4$. These new plateaus cannot be understood from
Landau quantization alone, and their origin is currently under
extensive theoretical study and considerable debate. In the next
section we review and compare the existing theoretical proposals for
these states, and discuss how to distinguish them experimentally. We
will also briefly mention theoretical work on edge states and
possible fractional quantum Hall state in graphene (which has not
been observed yet), and some other related issues, in section III.

\section{Spontaneous Symmetry Breaking and the New Quantum Hall States}

As discussed earlier and in Appendix A, fully occupied Landau levels
(LLs) of Dirac fermions that includes the spin and valley degeneracy
lead to integer QH states that belong to the sequence Eq.
(\ref{berry}). In the following for the sake of ease of discussion
we introduce a reduced filling factor $\nu_n$ for the
 highest LL (or valence LL, with index $n$) that is occupied by electrons:
\begin{equation}
\nu_n=\nu-4(n-1/2)\le 4.
\end{equation}
Obviously for the sequence (\ref{berry}), $\nu_n=4$ which reflects
the fact that the highest occupied LL is {\em fully} occupied, while
for the new QH states $\nu_n=1, 2$ or $3$, reflecting partial
occupation. In these new QH states the incompressibility or charge
gap cannot be due to LL spacing. The origin of the gap in these
cases is the subject of extensive theoretical work
recently\cite{nomura,gmd,alicea,ydm,abanin,gmss1,gmss2,herbut,ezawa,fl,ando}.
With few exception\cite{ando}, in these theories the gap has its
origin in some spontaneously broken symmetry of the system, and in
most
cases\cite{nomura,gmd,alicea,ydm,abanin,gmss1,gmss2,herbut,ezawa}
the symmetry breaking is driven by electron-electron interactions.
In the following we briefly review these theoretical works.

\subsection{Spontaneous breaking of SU(4) symmetry and quantum Hall ferromagnetism}

In this class of theories\cite{nomura,gmd,alicea,ydm,abanin}, an
(approximate) SU(4) symmetry that corresponds to the invariance of
electron-electron interaction under a rotation within the 4-fold
spin/valley internal space, is spontaneously broken. In the symmetry
breaking ground state, the system spontaneously picks $\nu_n$
orthogonal states in the SU(4) space, and fill them with electrons
in the $n$th LL:
\begin{equation}
|\Psi_0\rangle =\prod_{1\le\sigma\le
\nu_n}\prod_{k}c_{k,\sigma}^\dagger|0\rangle. \label{ground}
\end{equation}
Here $c^\dagger$ is the electron creation operator, $|0\rangle$ is
the vacuum state, $\sigma$ is the index of the internal state that
runs from 1 to 4, $k$ is an intra LL orbital index; for example in
the Landau gauge it is the wave vector along the plane wave
direction, while in the symmetric gauge it is the angular momentum
quantum number. Such a state can be shown to be an exact eigenstate
of the electron-electron interaction Hamiltonian in the presence of
full SU(4) symmetry, and very effectively gains exchange energy. For
a broad class of repulsive interactions, we expect $|\Psi_0\rangle$
to be the exact {\em ground state} of the system; this can be proven
rigorously for the case $\nu_n=1$ and short-range
repulsions\cite{ydm}. Any additional electron(s) added to the system
will then have to occupy the empty internal states, and lose
exchange energy; this results in additional energy cost, leading to
incompressibility of the system. The situation here is quite similar
to the ordinary single layer QH states in GaAs at odd integer
fillings, where electrons in the valence LL spontaneously magnetize
to lower exchange energy even in the absence of Zeeman splitting;
the spontaneously broken symmetry there is SU(2).\cite{sds,gm} Such
broken symmetry states are collectively known as quantum Hall
ferromagnets (QHF). In a recent numerical study\cite{cp}, QHF ground
states with large charge gaps have been observed for the graphene
SU(4) case.

In general symmetry breaking leads to low-energy collective modes.
Since SU(4) is a continuous symmetry, when broken spontaneously the
systems must support gapless collective modes. These neutral,
spin-wave like modes have been studied recently\cite{alicea,ydm},
and their full spectra have been determined exactly\cite{ydm}. Due
to the larger symmetry, we find that number of such modes is
$\nu_n(4-\nu_n)$, larger than the corresponding SU(2) case\cite{gm},
and is tied to the actual manner of symmetry breaking.\cite{ydm}

Perhaps more interesting is the appearance of topological solitons
in the order parameter called skyrmions as low-energy charge
excitations of the systems.\cite{sondhi,gm} Again due to the larger
symmetry, skyrmions come in more species in the SU(4) QHF\cite{ydm}
compared to the SU(2) QHF studied earlier.\cite{sondhi,gm} There is
also an important quantitative difference due to the Dirac nature of
the charge carriers in graphene. In a magnetic field, the eigen wave
functions of the two Dirac components are very different; they
actually each represent {\em different} LL wave functions of the
single component wave functions of non-relativistic
electrons\cite{nomura} (see also Appendix A). As a result the form
factor and thus effective electron-electron interaction in a given
LL is quite different for Dirac fermions.\cite{nomura,gmd} One of
the consequences of this is skyrmions are lowest energy charge
excitations not only in the lowest LL (which is the only case this
is true for of non-relativistic electrons\cite{wu}), but also in
some of the higher LLs\cite{alicea,ydm}; more specifically they are
lowest energy charge excitations for $|n|\le 3$.\cite{ydm} These
high LL skyrmions have been observed in a recent numerical
work.\cite{toke}

Small perturbations that explicitly break the SU(4) symmetry do
exist in graphene; they may fix the direction of the SU(4) order
parameter and open small gap(s) in the collective mode spectra. In
the following we discuss different types of such perturbations.

{\em Zeeman Splitting}. This is probably the most obvious one, which
lifts the spin degeneracy without affecting the valley degeneracy;
formally it reduces the SU(4) symmetry to a (valley) SU(2) symmetry.
Its effect is particularly important when $\nu_n=2$, in which case
it uniquely selects out a single ground state which is a valley
singlet but spin fully polarized in the valence LL, with or without
electron-electron interaction. In the absence of interaction the gap
of this state is precisely the Zeeman splitting, and in some
theories (to be discussed later) this is thought to be the sole
source of gap for $\nu=\pm 4$. Quantitatively, the Zeeman splitting
is of order 10 K, while electron-electron interaction scale
$e^2/\epsilon\ell \sim 100 - 1000K$ ($\ell=\sqrt{\hbar c/eB}$ is the
magnetic length); it thus qualifies as a small symmetry breaking
perturbation.

{\em Lattice Effect}. The SU(4) symmetry is exact in the continuum
limit $a \ll \ell$, where $a$ is lattice spacing. The lattice effect
breaks the SU(4) symmetry, or more precisely, the SU(2) symmetry
associated with valley symmetry.\cite{gmd,alicea} The strength of
the symmetry breaking is proportional to $a/\ell$. For the lowest LL
($n=0$), this introduces an Ising anisotropy in the valley SU(2)
space,\cite{alicea} while for $|n|>0$ an easy-plane (or XY)
anisotropy is introduced.\cite{gmd,alicea}

{\em Disorder}. Even if the disorder potential did not break any
symmetry, in general it lifts LL degeneracy and favors SU(4) singlet
ground states. In Ref. \onlinecite{nomura}, where the notion of QHF
was first introduced in the graphene context, a Stoner-like
criterion was derived for the development of QHF against disorder.
The situation was further analysed in Ref. \onlinecite{alicea},
where the authors came to the conclusion that the existing samples
may not be clean enough for QHF, and the system may actually be
SU(4) paramagnets; they argue that the combination of Zeeman and
symmetry breaking interactions can give rise the new QH states in
the paramagnetic regime.

On the other hand Abanin {\em et al.}\cite{abanin} pointed out an
interesting symmetry breaking effect of disorder potential, which
may {\em favor} ordering: the random potential due to strain in the
graphene lattice locally breaks the valley degeneracy in a random
manner; it thus acts like a random field along $\hat{z}$ direction
in the valley subspace. This forces the valley SU(2) order parameter
into the XY plane, and allows for algebraic long-range order at {\em
finite} temperature and a Kosterlitz-Thouless transition.

Before closing this subsection, we point out that many issues
similar to those discussed here have been studied in the context of
QHF in silicon\cite{arovas} and bilayer QH
systems\cite{gm,moon,ezawa03}. In particular the SU(4) symmetry was
used by Arovas {\em et al.}\cite{arovas} and Ezawa and
co-workers\cite{ezawa03} to organize the internal degrees of freedom
in silicon and bilayer systems respectively. It should be noted
however that graphene is a much better realization of the SU(4)
symmetry due to the small parameters that control the
symmetry-breaking perturbations. Such small parameters apparently do
not exist in silicon or bilayer systems; the origin of symmetry
breaking is well-understood in the latter case while it is much less
obvious in the former.

\subsection{Spontaneous mass generation and symmetry breaking}

In another set of theories,\cite{gmss1,gmss2,herbut,ezawa} it was
argued that the system spontaneously generates a mass for the Dirac
particles by breaking the translational symmetry, and the symmetry
breaking is driven by electron-electron interaction. Such
spontaneous mass generation was considered even before the actual
experimental realization of monlayer graphene.\cite{khvesh} In Ref.
\onlinecite{herbut}, it was argued that the ordering that generates
the mass could be either charge density way (CDW) or
antiferromagnetism (AFM), depending on the relative strength of
on-site and nearest neighbor electron-electron repulsion. The main
selling point of these theories is the fact that a Dirac mass would
lift the valley degeneracy in the $n=0$ LL {\em only}, but not for
$|n| \ge 1$ (see Appendix A for a discussion of this point).
Combining with Zeeman splitting, these theories would predict that
in the $n=0$ LL one would get QH states for all integer $\nu_n$, but
for $|n| \ge 1$ only even $\nu_n$ (and thus $\nu$) support QH
states. This is consistent with existing experimental
results\cite{zhang2} that the odd integer QH states observed thus
far, $\nu=\pm 1$, are in the $n=0$ LL.

\subsection{Electron-phonon interaction and spontaneously broken inversion symmetry}

Fuchs and Lederer\cite{fl} argued that in the presence of a magnetic
field, electron-phonon interaction can induce a spontaneous lattice
distortion similar to the Peierls effect, which breaks the inversion
symmetry of graphene. This spontaneously broken inversion symmetry
also generates a Dirac mass, although electron-electron interaction
only plays a minor role here. The consequences of this mass
generation are the same as the theories discussed in section IIB.

\subsection{Comparison of competing theories and experimental distinctions}

It should be obvious from the discussions above that the existing
theories all share a common theme that spontaneous symmetry breaking
plays a central role in stabilizing the new QH states observed in
Ref. \onlinecite{zhang2}. They are not entirely orthogonal in
details either. For examples, theories of subsections B and C both
rely on spontaneous mass generation; in the QHF theories described
in subsection A, the valley QHF for the $n=0$ LL {\em is} a CDW due
to the fact that wave functions of different valleys live in
opposite sublattices. In the following we focus on their differences
and how to distinguish among them experimentally.

Perhaps the first question to address is whether the origin of these
new QH states lies in electron-electron interaction or other
mechanism, like electron-phonon interaction. Experimentally, this
question can be settled by measuring the magnetic field ($B$)
dependence of the gap at {\em fixed} filling factors, in
sufficiently clean samples. If electron-electron interaction
dominates the gap $\Delta$, we expect $\Delta\propto
e^2/\epsilon\ell\propto\sqrt{B}$; such a behavior would be strong
indication of the electron-electron interaction mechanism. Also the
size of $\Delta$ is an indicator too; $e^2/\epsilon\ell$ is by far
the largest energy scale in the system other than LL spacing.

The biggest difference between the QHF theory and other theories
discussed in subsections B and C is it allows for QH states at all
integer fillings, including odd integers in $|n|\ge 1$ LLs, in
sufficiently clean samples. It would thus be highly desirable to
have higher quality samples and study if more QH states can be
found, including odd integers other than $\nu=\pm 1$; if so this
would lend strong support to QHF. In addition, due to the breaking
of a continuous symmetry in QHF, they support highly collective
neutral and charge excitations, which are not shared in other
theories. The neutral, spin-wave like modes may be detectable in
inelastic light scattering or optical absorption
experiments,\cite{iyengar,note} while charged skyrmions may be
detected in transport and other experiments;\cite{skyrexpt} in
particular, valley skyrmions have been seen in silicon
samples.\cite{mansour} Some of these experiments performed in
semiconductors to detect skyrmions can also be done in graphene.
Obviously evidence for these neutral and charged collective
excitations would also lend strong support to the QHF theory.

It is clear from the discussion above that higher quality (or
mobility) samples in which disorder effects are minimal are crucial
to further progress on the issues discussed in this section.

\section{Edge States, fractional quantum Hall states, and
other issues}

{\em Edge States}. Edge states play a key role in QH
transport.\cite{kanefisher} Due to the special lattice structure of
graphene and Dirac nature of the carriers, the edge states here have
certain features not shared
elsewhere.\cite{neto,abanin2,brey,fertig,hatsugai} Among many
peculiar properties, the edge states of the $\nu=0$ QH state (which
clearly has no counterpart elsewhere) are particularly interesting
-- they are argued to carry no charge current but finite spin
current in equilibrium.\cite{abanin2} There appear to be some
experimental evidence in support of this.\cite{abanin3} The
possibility of edge reconstruction\cite{reconstruct} has also been
discussed.\cite{neto}

{\em Possible Fractional Quantum Hall States}. Perhaps the most
eagerly anticipated next experimental breakthrough is the
observation of fractional Quantum Hall (FQH) States. Possible FQH
states have been discussed in a number of
papers,\cite{peres,ydm,apalkov,toke,toke2,goerbig} and some of them
may be QHF that break SU(4) symmetry as
well.\cite{ydm,toke,toke2,goerbig} Possible compressible, composite
fermion Fermi sea states have also been
discussed.\cite{khveshchenko,baskaran}

{\em Lattice Effects, Disorder, Localization, and Integer Quantum
Hall Transition.} Most of the papers mentioned above use the
continuum description of Dirac fermions, which properly describe
low-energy electronic state when the magnetic field strength is not
too strong. When the field gets so strong that $a/\ell$ is no longer
small, or when the Fermi energy becomes comparable to band width,
the continuum description breaks down and the lattice effect must be
taken into account. This issue has been addressed by studying the QH
physics on a honeycomb lattice
directly.\cite{ando,hatsugai,sheng,hk,bernevig} The lattice model
also allows for a straightforward numerical study of effects of
disorder, and the associated localization and integer QH transition
for non-interacting electrons problem.\cite{ando,sheng,hk,bd} In
particular, it was shown that disorder can split the critical states
carrying non-zero Chern numbers associated with the two degenerate
valleys in the absence of disorder; this offers yet another
mechanism for the new QH states discussed earlier that does not
invole any interaction. Most of the lattice work relies heavily on
the calculation of the Chern number, which have been used
extensively before in the study of both integer\cite{chern} and
fractional\cite{chern1} quantum Hall effects.

\section{Concluding remarks}

We hope the discussions above have made it clear that progress made
in the field of quantum Hall effect in graphene thus far has been
exciting, but the whole business is in its very early stage; many
important issues remain unresolved. As usual experimentalists hold
the key to further progress, and higher sample quality is most
crucial to such progress.

\acknowledgements The author is indebted to Sankar Das Sarma and
Allan MacDonald for fruitful collaboration and numerous stimulating
discussions on this subject, as well as useful comments and
suggestions on the manuscript. He also thanks Zhigang Jiang for
keeping him informed of experimental developments of quantum Hall
effect in graphene, and Shou-Cheng Zhang for useful discussions on
SU(N) symmetry. This work was supported by National Science
Foundation grant DMR-0225698.

\appendix
\section{Landau levels of Dirac particles}

In this Appendix we discuss the Landau levels (LLs) of Dirac
particles. Instead of repeating the theoretical treatment well
documented in literature, here we provide a heuristic discussion of
the LL energies and degeneracies, without actually writing down the
Dirac equation.

{\em Landau Level Energies.} The dispersion relation of relativistic
particles with mass $m$ is
\begin{equation}
E=\sqrt{m^2v_F^4+v_F^2p^2}.
\end{equation}
In the graphene context Fermi velocity $v_F$ plays the role a speed
of light, and the mass $m$ is either zero or extremely small. In the
presence of a magnetic field $B$, classically the particle exercises
cyclotron motion, and the semiclassical Bohr-Sommerfeld rule leads
to the quantization condition for the radii of such cyclotron
orbits:
\begin{equation}
\pi R_n^2=nhc/eB,
\end{equation}
i.e., they enclose an integer number of flux quanta. This dictates
the mechanical momentum of the particle to be
\begin{equation}
p_n=\sqrt{2n}\hbar/\ell,
\end{equation}
where $\ell=\sqrt{\hbar c/eB}$ is the magnetic length. Substituting
into (A1), we obtain the Landau level energies:
\begin{equation}
E_n=\pm\sqrt{m^2v_F^4+2n\hbar v_F^2|eB|/c},
\end{equation}
where the negative energy states correspond to the Dirac hole
states, or the valence band states in the graphene context.
Amazingly, this is the {\em exact} result! Normally energy levels
obtained from the Bohr-Sommerfeld quantization condition misses the
zero point energy; here the effect of zero point motion is
apparently compensated for by the Berry phase shift discussed in
section I.

{\em Landau Level Degeneracy}. Let us start by ignoring the physical
spin, which contributes a trivial factor of two. We do explicitly
include the two-fold valley degeneracy which is important for the
following discussion. We know that two major triumphs of the Dirac
theory are (i) the particles must have ``spin"-1/2 to be consistent
with Lorentz invariance; and (ii) the magnetic $g$-factor is exactly
2 for the ``spin" in the non-relativistic limit. In the graphene
context, this ``spin" is actually the 2-fold sublattice degree of
freedom associated with the hexagonal lattice, or a pseudospin;
combining with the two-fold valley degeneracy we get a four
component wave function for usual Dirac particles. We also know that
$g=2$ has a remarkable consequence in the non-relativistic limit:
the Zeeman splitting is $g\mu_BB=\hbar\omega_c$, exactly the same as
the Landau level spacing! This leads to a degeneracy of
pseudospin-up states in the $n$th LL with the pseudospin-down states
in the $n-1$th LL; as a consequence of this each energy level has
degeneracy $2N_\Phi$ (where $N_\Phi$ is the number of flux quanta in
the system), and the wave functions of each degenerate level mix the
$n$th and $n-1$th LL wave functions. This is true except for the
level corresponding to $n=0$th LL with pseudospin up, whose
degeneracy remains $N_\Phi$. Now putting back the factor of two
associated with physical spin, one immediately obtains the
quantization condition (\ref{berry}) for the Hall conductance of
filled LLs. Remarkably, the degeneracy discussed above, and thus the
condition (\ref{berry}), persist even in the extreme relativistic
limit $m=0$, to be discussed below.

{\em The Zero Mass Limit}. In the limit $m\rightarrow 0$, nothing
significant happens to the $n\ge 1$ LLs. However the two $n=0$ LLs
with energies $\pm mv_F^2$ merge together to form a single level,
now with degeneracy $4N_\Phi$ just like the other LLs (including
physical spin). Since half of the states of this level come from the
valence band, it is half-filled at zero doping; as a consequence the
$\nu=0$ state corresponds to a partially filled LL!

Another way to phrase the discussion above is a nonzero Dirac mass
$m$, either intrinsic or dynamically generated, lifts the valley
degeneracy of the $n=0$ LL, but does not affect the degeneracy of
other LLs. This fact plays a crucial role in the theories discussed
in sections IIB and IIC.


\begin{thebibliography}{1}

\bibitem{graphene_exp} K. S. Novoselov {\em et al.}, Science {\bf 306}, 666 (2004);
Y. Zhang {\em et al.}, Phys. Rev. Lett. {\bf 94}, 176803 (2005); C.
Berger {\em et al.}, J. Phys. Chem. B {\bf 108}, 19912 (2004).

\bibitem{novoselov} K. S. Novoselov, A. K. Geim, S. V. Morozov, D. Jiang, M. I. Katsnelson, I. V. Grigorieva,
S. V. Dubonos and A. A. Firsov, Nature {\bf 438}, 197 (2005).

\bibitem{zhang} Yuanbo Zhang, Yan-Wen Tan, Horst L. Stormer and Philip Kim, Nature {\bf 438}, 201 (2005).

\bibitem{zheng} Y. Zheng and T. Ando, Phys. Rev. B {\bf 65}, 245420
(2002).

\bibitem{gusynin} V. P. Gusynin and S. G. Sharapov, Phys. Rev. Lett. {\bf 95},
146801 (2005).

\bibitem{peres}  N. M. R. Peres, F. Guinea, and A. H. Castro Neto,
 Phys. Rev. B {\bf 73}, 125411
(2006).

\bibitem{semenoff} G. W. Semenoff, Phys. Rev. Lett. {\bf 53},
2015 (1988).

\bibitem{haldane} F. D. M. Haldane, Phys. Rev. Lett. {\bf 61},
2449 (1984).

\bibitem{ahm83} A. H. MacDonald, Phys. Rev. B {\bf 28}, 2235 (1983).

\bibitem{schakel} A. M. J. Schakel, Phys. Rev. D {\bf 43}, 1428 (1991).

\bibitem{boebinger} K. S. Novoselov, Z. Jiang, Y. Zhang, S. V. Morozov, H. L. Stormer, U. Zeitler, J. C. Maan,
G. S. Boebinger, P. Kim, and A. K. Geim, Science {\bf 315}, 1379
(2007).

\bibitem{zhang2} Y. Zhang, Z. Jiang, J. P. Small, M. S. Purewal, Y.-W. Tan, M. Fazlollahi, J. D. Chudow,
J. A. Jaszczak, H. L. Stormer, P. Kim, Phys. Rev. Lett. {\bf 96},
136806 (2006).

\bibitem{nomura} K. Nomura and A. H. MacDonald, Phys. Rev. Lett. {\bf 96},
256602 (2006).

\bibitem{gmd} M. O. Goerbig, R. Moessner, and B. Doucot, Phys. Rev. B {\bf 74}, 161407
(2006).

\bibitem{alicea} J. Alicea and M. P. A. Fisher, Phys. Rev. B {\bf 74},
075422 (2006).

\bibitem{ydm} Kun Yang, S. Das Sarma, and A. H. MacDonald, Phys. Rev. B {\bf 74}, 075423
(2006).

\bibitem{abanin} Dmitry A. Abanin, Patrick A. Lee, and Leonid S.
Levitov, cond-mat/0611062.

\bibitem{gmss1} V.P. Gusynin, V.A. Miransky, S.G. Sharapov, and I. A. Shovkovy, Phys. Rev. B {\bf 74},
195429 (2006).

\bibitem{gmss2} V.P. Gusynin, V.A. Miransky, S.G. Sharapov, and I. A. Shovkovy,
cond-mat/0612488.

\bibitem{herbut} I. F. Herbut, Phys. Rev. B {\bf 75}, 165411 (2007).

\bibitem{ezawa} M. Ezawa, cond-mat/0609612; cond-mat/0606084.

\bibitem{khvesh} D. V. Khveshchenko, Phys. Rev. Lett. {\bf 87}, 206401
(2001).

\bibitem{fl} Jean-Noël Fuchs and Pascal Lederer, Phys. Rev. Lett. {\bf 98},
016803 (2007); cond-mat/0612386.

\bibitem{ando} Mikito Koshino and Tsuneya Ando, Phys. Rev. B {\bf 75},
033412 (2007).

\bibitem{sds}  S. Das Sarma and A. Pinczuk, eds. {\em Perspectives in
Quantum Hall Effects}, John Wiley and Sons, New York, 1997.

\bibitem{gm} For reviews, see articles by S. M. Girvin and A. H. MacDonald, and by J. P.
Eisenstein in Ref. \onlinecite{sds}.

\bibitem{cp} Tapash Chakraborty and Pekka Pietilainen,
cond-mat/0703536.

\bibitem{sondhi} S. L. Sondhi, A. Karlhede, S. A. Kivelson, and E. H. Rezayi, Phys. Rev. B
{\bf 47}, 16419 (1993).

\bibitem{wu} X.-G. Wu and S.L. Sondhi, Phys. Rev. B {\bf 51}, 14725 (1995).

\bibitem{toke} Csaba Toke, Paul E. Lammert, Jainendra K. Jain, and Vincent H.
Crespi, Phys. Rev. B {\bf 74}, 235417 (2006).

\bibitem{arovas} D. P. Arovas, A. Karlhede and D. Lilliehook, Phys. Rev. B {\bf 59},
13147 (1999).

\bibitem{moon} K. Yang, K. Moon, L. Zheng, A. H. MacDonald, S. M. Girvin,
D. Yoshioka, and Shou-Cheng Zhang, Phys. Rev. Lett. {\bf 72}, 732
(1994); K. Moon, H. Mori, K. Yang, S. M. Girvin, A. H. MacDonald, L.
Zheng, D. Yoshioka and S.-C. Zhang, Phys. Rev. B {\bf 51}, 5138
(1995); K. Yang, K. Moon, L. Belkhir, H. Mori, S. M. Girvin, A. H.
MacDonald, L. Zheng and D. Yoshioka, Phys. Rev. B {\bf 54}, 11644
(1996).

\bibitem{ezawa03} K. Hasebe and Z. F. Ezawa, Phys. Rev. B
{\bf 66}, 155318 (2002); Z. F. Ezawa, G. Tsitsishvili, and K.
Hasebe, Phys. Rev. B {\bf 67}, 125314 (2003); Z. F. Ezawa and G.
Tsitsishvili, Phys. Rev. D {\bf 72}, 085002 (2005); G. Tsitsishvili
and Z. F. Ezawa, Phys. Rev. B {\bf 72}, 115306 (2005).

\bibitem{iyengar} A. Iyengar, Jianhui Wang, H. A. Fertig, and L.
Brey, cond-mat/0608364.

\bibitem{note} We note in passing that cyclotron resonance has been
oberved in graphene recently using optical probes:  Z. Jiang, E.A.
Henriksen, L.C. Tung, Y.-J. Wang, M.E. Schwartz, M.Y. Han, P. Kim,
and H.L. Stormer, cond-mat/0703822; R.S. Deacon, K-C. Chuang, R.
J.Nicholas, K.S. Novoselov, and A.K. Geim, cond-mat/0704041.

\bibitem{skyrexpt} S. E. Barrett, G. Dabbagh, L. N. Pfeiffer, K. W. West, and R. Tycko
Phys. Rev. Lett. {\bf 74}, 5112 (1995); A. Schmeller, J.P.
Eisenstein, L.N. Pfeiffer, and K.W. West, Phys. Rev. Lett. {\bf 75},
4290 (1995); E.H. Aifer, B.B. Goldberg, D.A. Broido, Phys. Rev.
Lett. {\bf 76}, 680 (1996); V. Bayot, E. Grivei, S. Melinte, M.B.
Santos, and M. Shayegan, Phys. Rev. Lett. {\bf 76}, 4584 (1996); V.
Bayot, E. Grivei, J.-M. Beuken, S. Melinte, and M. Shayegan, Phys.
Rev. Lett. {\bf 79}, 1718 (1997); D.R. Leadley, R.J. Nicholas, D.K.
Maude, A.N. Utjuzh, J.C. Portal, J.J. Harris, and C.T. Foxon, Phys.
Rev. Lett. {\bf 79}, 4246 (1997); J.L. Osborne, A.J. Shields, M.Y.
Simmons, N.R. Cooper, D.A. Ritchie, and M. Pepper, Phys. Rev. B {\bf
58}, R4227 (1998); S. Melinte, E. Grivei, V. Bayot, and M. Shayegan,
Phys. Rev. Lett. {\bf 82}, 2764 (1999); J.H. Smet, R.A. Deutschmann,
F. Ertl, W. der Wegschei, G. Abstreiter and K. von Klitzing, Phys.
Rev. Lett. {\bf 92}, 086802 (2004); P.G. Gervais, H.L. Stormer, D.C.
Tsui, P.L. Kuhns, W.G. Moulton, A.P. Reyes, L.N. Pfeiffer, K.W.
Baldwin, and K.W. West, Phys. Rev. Lett. {\bf 94}, 196803 (2005).

\bibitem{mansour} Y.P. Shkolnikov, S. Misra,
N.C. Bishop, E.P. De Poortere, and M. Shayegan, Phys. Rev. Lett.
{\bf 95}, 066809 (2005).

\bibitem{kanefisher} For a review, see C. L. Kane and M. P. A.
Fisher, in Ref. \onlinecite{sds}.

\bibitem{neto} A. H. Castro Neto, F. Guinea, and N. M. R. Peres, Phys. Rev. B {\bf 73}, 205408
(2006).

\bibitem{abanin2} Dmitry A. Abanin, Patrick A. Lee, and Leonid S.
Levitov, Phys. Rev. Lett. {\bf 96}, 176803 (2006).

\bibitem{brey}  L. Brey and H.A. Fertig, Phys. Rev. B {\bf 73}, 235411
(2006).

\bibitem{fertig}  H.A. Fertig and Luis Brey, Phys. Rev. Lett. {\bf 97}, 116805 (2006).

\bibitem{hatsugai} Y. Hatsugai, T. Fukui, and H. Aoki, Phys. Rev. B {\bf 74}, 205414 (2006).

\bibitem{abanin3} Dmitry A. Abanin, Kostya S. Novoselov, Uli Zeitler, Patrick A. Lee, Andre K. Geim, and Leonid S.
Levitov, cond-mat/0702125.

\bibitem{reconstruct} A.~H. MacDonald, S.~R.~E. Yang, and M.~D. Johnson, Aus.
J. Phys. {\bf 46}, 345 (1993); C. de C. Chamon and X.-G. Wen, {\em
ibid} {\bf 49}, 8227 (1994); X. Wan, K. Yang, and E.~H. Rezayi,
Phys. Rev. Lett. {\bf 88}, 056802 (2002); X. Wan, E.~H. Rezayi, and
K. Yang, Phys. Rev. B {\bf 68}, 125307 (2003); K. Yang, Phys. Rev.
Lett. {\bf 91}, 036802 (2003).

\bibitem{apalkov} V. M. Apalkov and T. Chakraborty, Phys. Rev. Lett. {\bf 97}, 126801 (2006).

\bibitem{toke2} Csaba Toke and Jainendra K. Jain, cond-mat/0701026.

\bibitem{goerbig} M. O. Goerbig and N. Regnault, cond-mat/0701661.

\bibitem{khveshchenko} D. V. Khveshchenko, Phys. Rev. B {\bf 75}, 153405 (2007).

\bibitem{baskaran} G. Baskaran, cond-mat/0702420.

\bibitem{sheng} D. N. Sheng, L. Sheng, and Z. Y. Weng, Phys. Rev. B {\bf 73}, 233406 (2006).

\bibitem{hk} Yasumasa Hasegawa and Mahito Kohmoto, Phys. Rev. B 74, 155415
(2006).

\bibitem{bernevig} B. Andrei Bernevig, Taylor L. Hughes, Han-Dong Chen, Congjun Wu, and Shou-Cheng
Zhang, Int. J. of Mod. Phys. B {\bf 20}, 3257 (2006).

\bibitem{bd} C. P. Burgess and B. P. Dolan, cond-mat/0612269.

\bibitem{chern} D.~J. Thouless, M. Kohomoto, M.~P. Nightingale,
and M. den Nijs, Phys. Rev. Lett. {\bf 49},  405 (1982); Q. Niu,
D.~J. Thouless and Y.~S. Wu, Phys. Rev. B {\bf 31}, 3372 (1985);
D.~P. Arovas {\em et. al.,} Phys. Rev. Lett. {\bf 60}, 619 (1988);
Y. Huo, and R.~N. Bhatt, Phys. Rev. Lett. {\bf 68}, 1375 (1992);
D.~N. Sheng and  Z.~Y. Weng, Phys. Rev. Lett. {\bf 75} 2388 (1995);
Phys. Rev. B{\bf 54}, R11070 (1996); K. Yang and R.~N. Bhatt, Phys.
Rev. Lett. {\bf 76}, 1316 (1996); Phys. Rev. B {\bf 59}, 8144
(1999); Phys. Rev. B {\bf 55}, R1922 (1997); R.~N. Bhatt and X. Wan,
Pramana-J. Phys. {\bf 58}, 271 (2002); Qinghong Cui, Xin Wan, and
Kun Yang, Phys. Rev. B {\bf 70}, 094506 (2004).

\bibitem{chern1} D. N. Sheng, Xin Wan, E. H. Rezayi, Kun Yang, R. N. Bhatt, and F. D. M.
Haldane, Phys. Rev. Lett. {\bf 90}, 256802 (2003);  D. N. Sheng,
Leon Balents, and Ziqiang Wang, Phys. Rev. Lett. {\bf 91}, 116802
(2003); Xin Wan, D.N. Sheng, E.H. Rezayi, Kun Yang, R.N. Bhatt, and
F.D.M. Haldane, Phys. Rev. B {\bf 72}, 075325 (2005).

\end{thebibliography}
\end{document}